# Human-Augmented Reality Interaction in Rebar Inspection

Mahsa Sanei[1], Fernando Moreu[1, *]

[1] Civil, Construction, and Environmental Engineering Department, University of New Mexico, Albuquerque, NM, 87131, USA

[*] Corresponding Author: Centennial Engineering Center 3051, 210 University Blvd, NE, MSC01 1070, Albuquerque, NM 87131. E-mail: fmoreu@unm.edu.

**Abstract**

Rebar inspection in reinforced concrete construction requires sustained awkward postures and complex mental mapping of two-dimensional drawings onto three-dimensional assemblies. This study evaluated an Augmented Reality (AR)-assisted rebar inspection system deployed on Microsoft HoloLens 2 through a within-subjects experiment with 30 participants. Full-body kinematics were recorded using a motion capture system at 100 Hz while participants performed traditional and AR-assisted spacing inspection. AR reduced mean trunk flexion by 30.8%, mean neck flexion by 32.8%, and task completion time by 67.7%. Walking distance and hand-path length each decreased by over 50%. NASA Task Load Index scores decreased by 45.6% overall, with the largest reduction in physical demand. Inspection accuracy was maintained across conditions. The System Usability Scale yielded a mean score of 76.1 with 83% of participants rating the system acceptable. These results provide convergent objective and subjective evidence that AR-assisted inspection reduces ergonomic risk and perceived workload maintaining inspection quality.

1. ## Introduction

Quality assurance in reinforced concrete construction depends on accurate rebar placement verification. Non-compliant spacing or positioning can compromise structural integrity and long-term durability [1], [2]. Traditional rebar inspection relies on manual measurements using tape measures, visual interpretation of two-dimensional drawings, and inspector judgment to verify complex three-dimensional rebar assemblies against design specifications [3]. This conventional approach presents challenges that impact both inspector well-being and inspection quality. First, the physical demands are substantial: inspectors must adopt and maintain awkward postures including trunk flexion, overhead reaching, and prolonged kneeling or squatting in congested reinforcement cages [4], [5]. Second, significant cognitive load arises from the mental transformation required to map 2D construction drawings onto 3D physical reality while simultaneously tracking multiple spacing requirements across horizontal and vertical directions [6]. Third, the sequential nature of manual measurement such as moving between reference points, recording measurements, comparing against specifications creates time inefficiencies that compound under project schedule pressures [7]. Fourth, human factors including inspector fatigue, environmental conditions, and task monotony introduce quality variability and increase error potential [8].

Augmented reality (AR) technology offers potential to address these challenges by overlaying digital information directly onto the physical workspace through head-mounted displays or mobile devices, eliminating cognitive load and enabling real-time, in-situ guidance [9]. Recent advances in AR hardware, particularly the Microsoft HoloLens 2 (HL2) with improved field of view, hand tracking, and spatial understanding capabilities, have enabled more practical construction applications [10], [11], [12]. Simultaneously, progress in depth sensing technologies including Azure Kinect and time-of-flight cameras has enabled automated geometric capture and analysis of as-built conditions [13], [14]. The integration of

Building Information Modeling (BIM) with AR visualization creates opportunities for automated deviation detection and correction guidance that could fundamentally transform inspection workflows [15], [16].

Despite growing interest in construction AR applications, a critical gap exists in empirical human factors evaluation. Many studies have demonstrated technical feasibility and measured task performance metrics such as accuracy and completion time [17], [18], however, systematic investigation of AR's impact on inspector physical workload, biomechanical stress patterns, cognitive demand, and technology acceptance remains limited [19], [20], [21]. This gap is particularly significant given that construction work exhibits among the highest rates of work-related musculoskeletal disorders (MSDs) across all industries, with inspection tasks involving sustained awkward postures [22], [23], [24]. Furthermore, technology adoption in construction depends not only on technical performance but critically on worker acceptance, perceived usability, and demonstrated ergonomic benefits [25], [26]. Therefore, the construction industry needs an empirical foundation to make informed technology adoption decisions, using comprehensive human factors evidence addressing physical, cognitive, and usability dimensions simultaneously.

## 2. Literature Review

### 2.1. AR Applications in Construction Inspection

Augmented reality applications in construction have advanced from visualizations to field-ready systems addressing specific inspection and quality control challenges. Recent systematic reviews have documented AR's expanding adoption across construction lifecycle phases, with inspection and quality control identified as primary application domains. Nassereddine et al. [27] analyzed use-cases, benefits, and obstacles of AR in construction through industry surveys and literature review, finding that AR applications clustered into three categories: decision support, task assistance, and situation prompting, with inspection tasks falling primarily under task assistance. The main barriers to AR adoption are high cost, immature technology, lack of standards, and unclear value, but many professionals still believe AR has strong future potential.

Several studies have demonstrated AR's capacity to improve deviation detection and quality verification through automated BIM-to-as-built comparison specifically in construction inspection. Chi et al. [28] developed a system integrating terrestrial laser scanning with HL-based AR for rebar inspection. It automatically compared point clouds with BIM geometry to detect spacing and position errors, visualized results in AR, and generated rework instructions. The system removed manual measurements but focused on technical feasibility rather than human factors. Similarly, Chalhoub et al. [29] evaluated AR for model reconciliation using HL. AR users reliably detected large deviations (>50 mm) and missing elements, but small deviations (<25 mm) were often missed and false positives occurred, indicating AR is better suited for qualitative checks than precise measurements. More recently, Tao et al. [30] proposed and evaluated mesh-to-mesh comparison methods for mixed reality–based Mechanical, Electrical, and Plumbing (MEP) progress monitoring using HL2 depth data, analyzing trade-offs between on-device processing speed and cloud-based accuracy.

AR has been also applied to different quality control scenarios. Wang et al. [31] developed an AR-assisted building assembly guidance system that provided real-time visualization of design information overlaid on physical components, reducing assembly errors and improving efficiency. May et al. [32] developed and evaluated a BIM-based AR defect management system on HL that integrated defect identification, documentation, and tracking. In a study with 36 participants, the AR system outperformed paper-based methods in task time, defect detection accuracy, and user preference, demonstrating clear advantages for inspection workflows. Tan et al. [33] integrated YOLOv5 computer vision into an AR-BIM defect inspection system, achieving centimeter-level precision and 78.6% efficiency improvement over manual methods through automated defect detection and classification deployed on HL2. Mirshokraei et al. [34] developed a web-based BIM–AR quality management system for structural elements that generated AR inspection checklists from BIM models. The system addressed key quality management issues, including poor information management, unclear quality scope, and weak stakeholder communication. However, these studies focused on task performance metrics without objective measurement of physical ergonomic impacts.

### 2.2. Ergonomic Risks and Physical Demands in Construction Inspection Work

Construction work imposes substantial physical demands that cause high rates of MSD. According to the U.S. Bureau of Labor Statistics [35], construction workers experience MSDs at rates 40% higher than the all-industry average, with back injuries, shoulder strains, and neck disorders representing the most prevalent injury types. Umer et al. [36] conducted a systematic review of 35 studies on musculoskeletal symptoms in construction, finding that more than 50% of construction workers suffer from lower back symptoms, with rebar workers identified as particularly susceptible due to the postural demands of their work. Similarly, Anwer et al. [37] reviewed 20 years of literature on work-related musculoskeletal disorders (WMSD) in construction, confirming that physical risk factors including awkward postures, repetitive movements, and prolonged static loading were strongly associated with WMSDs, with back disorders accounting for 41.7% of construction workers' days away from work.

Research on rebar work has quantified biomechanical risk factors for WMSDs. Umer et al. [38] used surface electromyography and inertial motion sensors to measure trunk kinematics during simulated rebar tying across three postures (stooping, squatting, kneeling), finding that all postures involved trunk inclination exceeding the ISO 11226 recommended 60° threshold for static work. In a subsequent study, Umer et al. [39] showed that a low-cost stool reduced trunk muscle activity and discomfort during rebar tying, demonstrating that task-level postural changes can lower WMSD risk. While these studies focused on rebar installation rather than inspection, the work highlights workers' biomechanical vulnerability and the benefits of reducing trunk flexion.

Objective biomechanical assessment quantifies physical demands and validates intervention effectiveness. Motion capture has become a key tool for continuous ergonomic monitoring in construction and industry. A 2025 systematic review by Salisu et al. [40] identified three primary motion capture modalities for ergonomic risk assessment: optical marker-based systems (e.g., Vicon, OptiTrack), inertial measurement unit (IMU) sensors, and markerless computer vision systems. Marker-based systems remain the accuracy gold standard with joint angle errors of 2-4° but require controlled environments and marker placement. IMU-based wearable sensors are portable and enable real-time field monitoring. Markerless systems are low-burden and scalable but sensitive to occlusion, clothing, and environment.

Recent studies have applied motion capture to construction ergonomics. Chen et al. [41] developed a deep learning–based 3D pose estimation model that extracts skeletal data from monocular video to compute Rapid Entire Body Assessment (REBA) scores in real time. Their approach achieved high accuracy in controlled testing but had challenges in complex field environments with occlusion and lighting variations. Zhou et al. [42] used Microsoft Kinect v2 depth sensors with Rapid Upper Limb Assessment (RULA) scoring for real-time posture analysis during maintenance and assembly, demonstrating feasibility of depth-based ergonomic monitoring. Multiple studies have also validated IMU-based wearable sensors for construction worker ergonomic assessment. Yan et al. [43] developed a wearable IMU system using helmet and back sensors to detect hazardous trunk and neck postures, triggering alerts when ISO 11226 limits were exceeded. Yan et al. [44] extended this to rebar ironworkers, using IMUs with Ovako posture scoring to create personalized risk profiles and targeted interventions.

Despite extensive research on construction ergonomics and motion capture methods, a key gap remains: these techniques have not evaluated AR-assisted inspection as an ergonomic intervention. Marklin et al. [45] tested HL and RealWear HMT-1 with utility workers, finding similar neck muscle activity but lower blink rates with HL, suggesting potential eye strain. Kuber and Rashedi [46] reviewed studies showing neck and shoulder demands vary with headset design and target placement. While AR can improve inspection efficiency, it is still unknown whether AR-assisted inspection reduces biomechanical load, including trunk and neck flexion, movement patterns, and time-in-task.

### 2.3. Cognitive Workload and Usability Evaluation of Augmented Reality Systems

Assessing the cognitive demands and usability of AR systems is crucial, since adoption depends on user acceptance, manageable workload, and overall experience. The NASA Task Load Index (NASA-TLX) [47], originally developed for aviation, is now the standard subjective workload tool, measuring mental demand, physical demand, temporal demand, performance, effort, and frustration. Several recent studies have specifically applied NASA-TLX to evaluate AR systems in construction contexts. Abbas et al. [21] compared mobile AR-assisted rebar inspection with traditional paper-based methods, measuring NASA-TLX, situational awareness (SART), and task performance. They found that AR reduced some cognitive demands but could also cause overload due to information superimposition, and its limited field of view sometimes lowered situational awareness compared to paper. Jia et al. [48] evaluated AR head-mounted displays for construction assembly tasks using NASA-TLX combined with EEG-based mental workload measurement, finding complex trade-offs between task performance improvements and increased cognitive demands related to visual attention and information processing. Wu et al. [49] tested an ergonomics-focused optical see-through AR system with 40 participants and found lower NASA-TLX scores than paper instructions across all subscales, along with improved accuracy and completion time, showing that reduced cognitive load translated into real performance gains. In industrial quality inspection, Sauer et al. [50] used NASA-TLX to evaluate a head-mounted AR system that visualized defects directly on products. In a within-subjects study with 35 participants, AR significantly reduced cognitive load, especially mental demand and effort and improved task performance, particularly for more difficult inspections.

The System Usability Scale (SUS), developed by Brooke [51], provides a simple, reliable method for evaluating perceived usability through 10 items yielding a composite score from 0 to 100. SUS has been extensively validated across digital interfaces and achieves high reliability. Recent research has applied SUS to evaluate AR construction applications. May et al. [32] assessed their BIM-AR defect management system using both SUS and NASA-TLX with 36 participants, finding that the AR system achieved a mean SUS score of 73.1, indicating "good" usability in Bangor et al.'s [52] adjective rating classification scheme, while simultaneously demonstrating significantly lower NASA-TLX workload scores than paper-based inspection methods. Their study exemplified comprehensive human factors evaluation by combining SUS, cognitive workload, task performance (completion time, accuracy), and user preference measures to provide convergent evidence of AR's benefits.

This study extended prior AR construction research by providing simultaneous objective and subjective human factors evaluation. This is critical in construction inspection, where cognitive and physical demands are closely linked. The evaluation integrates objective biomechanical analysis of body posture and movement using motion capture system with validated subjective measures, including NASA-TLX for workload and SUS for usability. The research objectives are to: (1) Quantify biomechanical differences between traditional and AR-assisted inspection, including trunk flexion and neck extension, (2) Assess task performance changes, including completion time, walking distance, and movement efficiency, (3) Evaluate multidimensional subjective workload across mental, physical, temporal demand, effort, performance, and frustration, and (4) Determine system usability and factors affecting technology acceptance.

### 3. Research Methodology

This study employed a within-subjects experimental design where each participant completed rebar spacing inspection task under two conditions: (1) Traditional manual inspection using tape measure and (2) AR-assisted inspection. 30 participants performed both inspection conditions following by filling out two surveys (NASA-TLX and SUS). Statistical analysis were conducted for collected data from motion capture system and two surveys.

*3.1. Rebar Inspection System AR Application Design*

The AR-based rebar inspection application is developed using Unity 2020.3.28 and MRTK 2.8. This system implements a distributed client-server architecture that enables real-time visualization of rebar positioning corrections on Microsoft HL2. The workflow begins when the user establishes a TCP network connection between the HoloLens device and a MATLAB-based server by entering the server's IP address through the AR interface. The server-side processing pipeline performs tasks including point cloud processing, automated rebar detection, comparison with BIM specifications, and generation of optimized correction sequences. This architectural separation enables efficient deployment on the HL platform while leveraging MATLAB's computational capabilities for complex geometric analysis. Figure 1 shows the hologram of the as-built model overlaid on the real rebar using QR code and the menu where human inspector can see the correction instructions.

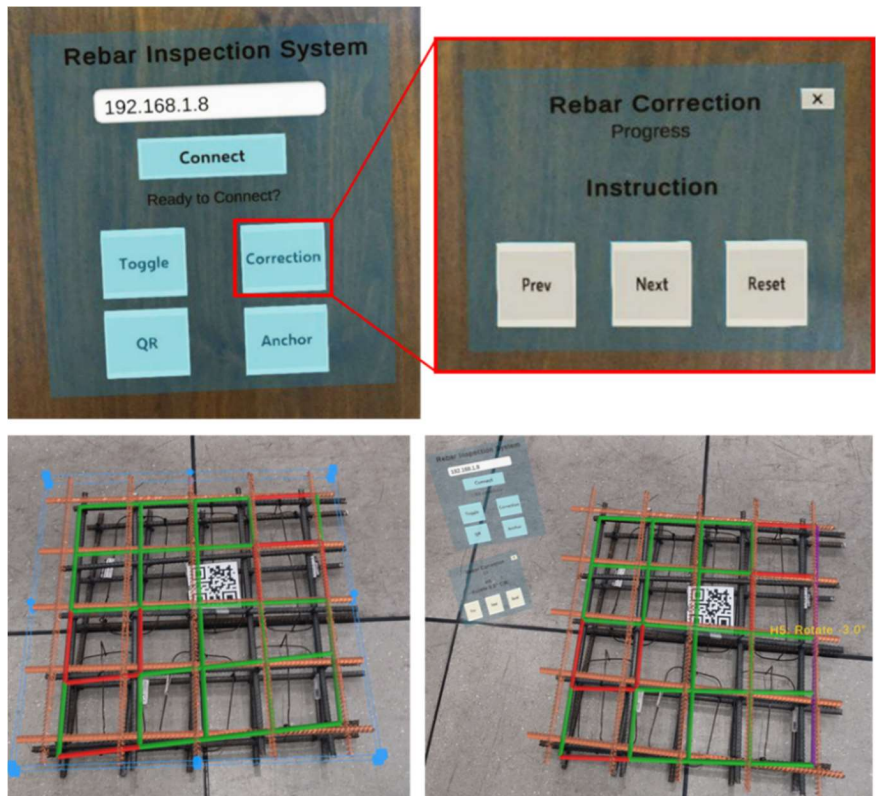

Figure 1. Rebar spacing inspection AR menu and hologram of rebar correction

A QR code-based anchoring procedure is used to establish coordinate system alignment between the physical environment and digital models. Users can save model anchored positions to HL storage. Following successful anchoring, inspectors can click on "correction" button to open up the Rebar Correction Menu and navigate through the optimized correction sequence. The interface presents a progress indicator showing current position, instructions describing specific adjustments, and visual overlays highlighting the target rebar element in the physical scene. The AR interface provides intuitive navigation controls including Previous/Next buttons for sequential progression and a Reset function for workflow restart if field conditions change. Table 1 shows the description of the buttons in the developed AR interface.

Table 1. AR rebar inspection menu function description

| Menu | UI Component | Type | Functional Description | System-Level Operation |
|------|--------------|------|------------------------|------------------------|
| Rebar Inspection System (Main menu) | Server IP Input | Text Input Field | User enters the IPv4 address of the MATLAB processing server | Used to initialize TCP client socket and establish bidirectional communication |
| | Connect | Action Button | Initiates connection to backend server | Creates TCP socket session; requests and loads 3D rebar model into Unity scene |
| | Toggle Manipulation | Toggle Button | Enables/Disables manual manipulation of 3D model | Activates/deactivates bounding box collider and transformation handles |
| | Correction | Action Button | Opens correction workflow interface | Loads correction UI canvas and initializes step-based inspection routine |
| | QR Detection | Action Button | Activates QR-based spatial registration | Starts HoloLens camera stream; performs QR detection; computes pose and aligns 3D model to physical object |
| | Anchor | Action Button | Saves spatial alignment | Stores world anchor to preserve hologram pose in spatial coordinate system |
| Rebar Correction Menu | Progress Indicator | Status Display | Displays current correction step and total steps | Updates dynamically from correction state manager |
| | Instruction Panel | Text Display | Displays corrective action instructions | Retrieves step-specific instruction from correction database |
| | Previous | Navigation Button | Returns to previous correction step | Decrements step index and refreshes UI state |
| | Next | Navigation Button | Advances to next correction step | Increments step index and updates model visualization |
| | Reset | Action Button | Restarts correction workflow | Resets step index and clears temporary correction states |

*3.2. Motion Capture Objective Assessment*

Full-body kinematics were recorded using a 10-camera Vicon Valkyrie motion capture system [53] operating at 100 Hz sampling frequency. The global coordinate system was defined with the origin at the center of the rebar assembly, X-axis aligned with the long dimension, Y-axis vertical, and Z-axis completing the right-handed coordinate system. For this analysis, marker data from four key body segments were extracted: the upper back, the head, and bilateral hand markers. These segments were selected based on their relevance to the ergonomic risk factors of interest (trunk flexion, neck flexion, and hand movement patterns). Figure 2 shows the Vicon motion capture system and object attached to the inspector's body for tracking their movement.

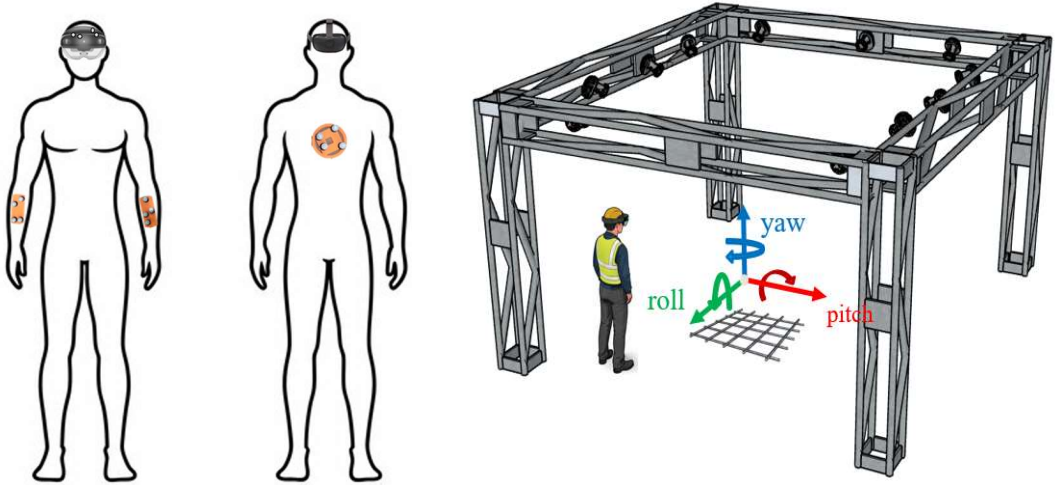

Figure 2. Motion capture system for inspector's body tracking

Processed motion capture data were imported into a MATLAB analysis pipeline for calculation of biomechanical variables. The following categories of variables were computed based on their relevance to known ergonomic risk factors in construction work: (1) Movement efficiency metrics: Total walking distance was computed as the cumulative Euclidean distance traveled by the back marker throughout the task. Task duration was extracted from synchronized motion capture timestamps, (2) Postural variables: Trunk flexion angle and neck flexion angle. Trunk flexion angle was defined as the inclination of the upper-back rigid-body longitudinal axis relative to the global vertical. Let $\hat{e}_{trunk}$ denote the unit vector along the cranial–caudal axis of the upper-back rigid body extracted from Vicon orientation data, and $\hat{e}_Z$ the global vertical unit vector. Trunk flexion was computed as equation (1). Neck flexion angle was defined as the relative angular deviation between the head and upper-back rigid bodies. Let $\hat{e}_{hea}$ represent the unit vector along the cranial–caudal axis of the head rigid body. Neck flexion was computed as equation (2).

$$\theta_{trunk} = \arccos(\hat{e}_{trunk} \cdot \hat{e}_Z) \tag{1}$$

$$\theta_{neck} = \arccos(\hat{e}_{head} \cdot \hat{e}_{trunk}) \tag{2}$$

Where $\theta_{trunk} = 0°$ corresponds to a fully upright posture and $\theta_{neck} = 0°$ shows the head is aligned with the trunk. Figure 3 shows the trunk and neck flexion angle definition.

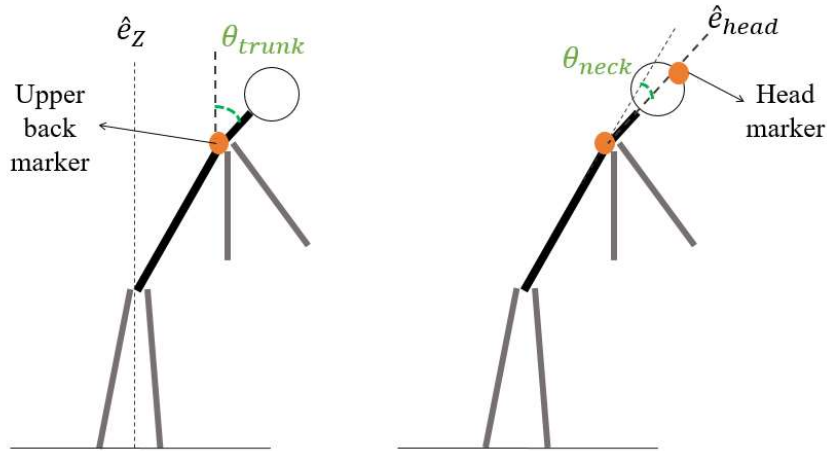

Figure 3. Trunk and neck flexion angle definition

Ergonomic risk during rebar inspection is influenced by joint flexion angle and posture duration. According to ISO 11226 [54], trunk inclination below 20° is classified as acceptable regardless of duration, angles between 20° and 60° are conditionally acceptable subject to a maximum acceptable holding time (MAHT) derived from the Rohmert endurance model [55], and any posture exceeding 60° is not recommended at any duration. For the neck, ISO 11226 defines a binary threshold: relative neck flexion (β − α) exceeding 25° is not recommended, while absolute head inclination between 25° and 85° is conditionally acceptable with time-dependent limits. The REBA [56] and RULA [57] observational tools support these boundaries, with REBA assigning the highest trunk score (4) at flexion beyond 60° and the highest neck score (2) beyond 20°, while RULA provides a finer 10° neck transition threshold. Other studies support these thresholds: Hoogendoorn et al. [6] found a relative risk of 1.5 for low back pain when trunk flexion exceeded 60° for more than 5% of working time, and Ariëns et al. [58] demonstrated increased neck pain risk with sustained neck flexion beyond 20°. Field studies of rebar workers confirm that these thresholds are routinely exceeded. Umer et al. [36] showed that stooping, kneeling, and squatting rebar tying postures all produce trunk inclination above 60°, while Yan et al. [43] implemented the ISO 11226 MAHT curves in a wearable IMU-based alert system and found that personalized thresholds improved ergonomic outcomes over generic limits. Figure 4 presents the angle–time risk map created from these standards.

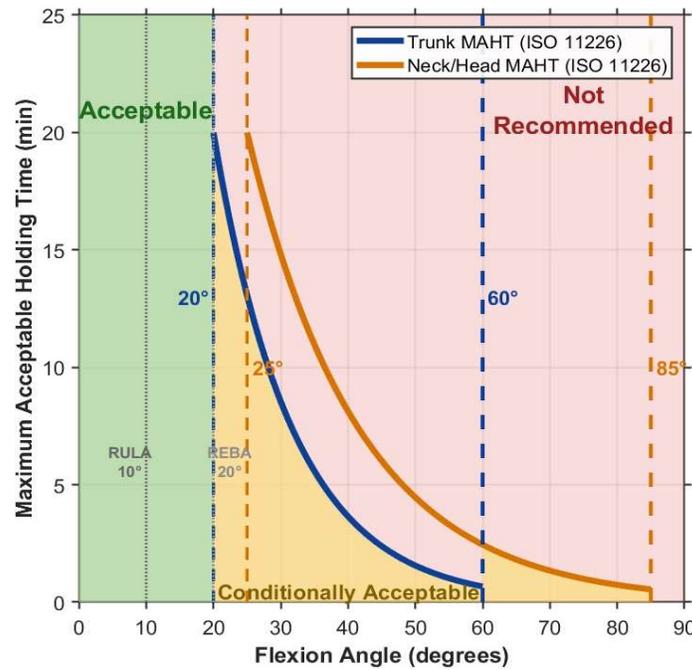

Figure 4. Neck and trunk flexion angle vs. time risk zones ISO11226

These definitions capture gross segmental inclination and inter-segmental deviation, respectively, and differ from clinical trunk flexion measures that reference the pelvis. Because no pelvis marker was used in the present setup, $\theta_{trunk}$ reflects thoracolumbar inclination from the global vertical rather than isolated lumbar flexion. Similarly, $\theta_{neck}$ represents the head-to-trunk relative angle rather than isolated cervical vertebral motion. The ISO 11226 thresholds cited in this study were originally defined using similar gross inclination measures (trunk inclination from vertical, head inclination relative to trunk), making the present definitions compatible with those standards

### 3.3. Statistical Analysis

All statistical analyses were performed using MATLAB R2024a [59]. Normality of paired differences was assessed using the Shapiro-Wilk test [60]. For normally distributed data (p > 0.05), paired t-tests were conducted to compare AR versus traditional conditions; for non-normally distributed data, Wilcoxon signed-rank tests were applied. Effect sizes were calculated using Cohen's d for parametric tests and rank-biserial correlation ($r = Z/\sqrt{n}$) for non-parametric tests, following conventions where $|d| > 0.8$ indicates large effects, 0.5–0.8 medium effects, and 0.2–0.5 small effects [61]. Statistical significance was set at $p < 0.05$ (two-tailed), with 95% confidence intervals reported for mean differences.

### 3.4. Subjective Assessment Instruments

Participants completed the NASA-TLX following each inspection condition, rating each subscale on a scale from 0 (very low) to 100 (very high) as shown in Table 2. Scores were compared using paired t-tests or Wilcoxon tests as appropriate. The Performance dimension was reverse-scored (100 − raw score) prior to analysis to ensure consistent directionality across all subscales, where higher values indicate worse outcomes.

Table 2. NASA-TLX questionnaire

| # | Dimension | Description |
|---|---|---|
| 1 | Mental Demand | How mentally demanding was the task? |
| 2 | Physical Demand | How physically demanding was the task? |
| 3 | Temporal Demand | How hurried or rushed was the pace of the task? |
| 4 | Performance | How successful were you in accomplishing what you were asked to do? (0 = Perfect, 100 = Failure) |
| 5 | Effort | How hard did you have to work (mentally and physically) to accomplish your level of performance? |
| 6 | Frustration Level | How insecure, discouraged, irritated, stressed, or annoyed were you? |

System usability was evaluated using the SUS questionnaire (Table 3) that provides a valid measure of subjective usability. The SUS was administered only after the AR-assisted inspection condition, as it is specifically designed to assess technology system usability. Each item is rated on a 5-point scale from 1 (Strongly Disagree) to 5 (Strongly Agree).

Table 3. SUS questionnaire

| # | Statement |
|---|---|
| 1 | I think that I would like to use this AR system frequently. |
| 2 | I found the AR system unnecessarily complex. |
| 3 | I thought the AR system was easy to use. |
| 4 | I think that I would need the support of a technical person to be able to use this AR system. |
| 5 | I found the various functions in this AR system were well integrated. |
| 6 | I thought there was too much inconsistency in this AR system. |
| 7 | I would imagine that most people would learn to use this AR system very quickly. |
| 8 | I found the AR system very cumbersome to use. |
| 9 | I felt very confident using the AR system. |
| 10 | I needed to learn a lot of things before I could get going with this AR system. |

## 4. Experimental Design

### 4.1. Participants

The experimental protocol received approval from the University of New Mexico Institutional Review Board (IRB Protocol # 2510239462). All participants provided written informed consent after receiving detailed explanation of study procedures, potential risks, and data usage. Participants were informed of their right to withdraw at any time without penalty. The consent process emphasized that motion capture data would be de-identified and stored securely, with access limited to the research team. All participants completed both experimental conditions with valid motion capture data. Table 4 shows the participants' characteristics.

Table 4. Participant characteristics (N = 30)

| Category | Variable | Value |
|---|---|---|
| Age | Mean ± SD | 28.8 ± 5.9 years |
| | Range | 20 – 41 years |
| Gender | Male | 17 (57%) |
| | Female | 13 (43%) |
| Participant role | Graduate students | 18 (60%) |
| | Professionals | 9 (30%) |
| | Undergraduate students | 3 (10%) |
| Inspection familiarity | Mean score | 3.6 ± 1.7 years |
| | Scale description | 1 = Aware, 2 = hands-on, 3 = Info only, 4 = Heard of it, 5 = No idea |
| AR Experience | Mean score | 2.8 / 5 |
| | Scale description | 1 = Own device/frequent use, 2 = Aware/hands-on, 3 = Info only, 4 = Heard of it, 5 = No idea |

### 4.2. Experiment Specimen

The assembly consisted of 5 horizontal and 5 vertical reinforcing bars (No. 5, 15.875 mm diameter) arranged in two orthogonal layers with design spacing of 127 mm center to center. Intentional spacing violations were introduced at predetermined locations, with deviations ranging from 15 mm to 30 mm from the design spacing.

### 4.3. Experimental Tasks

The experimental task involved inspecting a standardized rebar assembly with two different methods as shown in Figure 5. In the traditional manual inspection conditions, participants received a paper copy of the construction drawings showing the design rebar layout, specified spacing dimensions, and tolerance requirements. They were provided with a standard measuring tape and clipboard for recording observations. Participants were instructed to systematically inspect the rebar assembly, identify all spacing violations exceeding the ±12.7 mm tolerance, record the location and magnitude of each violation, and propose corrections to bring the assembly into compliance. No specific inspection sequence was mandated, allowing participants to follow their natural inspection strategies.

In the AR-assisted inspection condition, participants used HL2 displaying the AR inspection application. After a brief training period (approximately 10 minutes) familiarizing participants with hand gestures, they performed the same inspection task. The AR system automatically highlighted spacing violations with color-coded overlays visible through the headset's transparent display. Participants could select "correction" button to view detailed correction instructions including movement direction and distance.

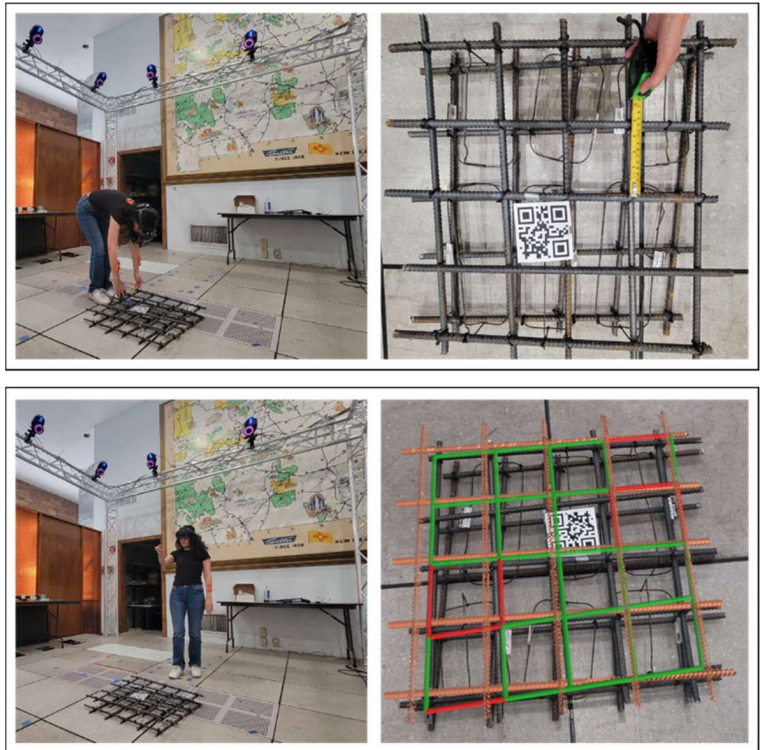

Figure 5. Rebar spacing inspection under two conditions: Traditional and AR-assisted

In both methods, participants physically adjusted rebar positions. In Traditional condition, participants manually identified violations, determined corrections, and physically moved rebars to achieve compliance.

## 5. Results

### 5.1. Inspection Accuracy

Inspection accuracy was defined as the post-correction compliance rate: the percentage of 40 rebar spacings (20 vertical, 20 horizontal) that fell within the ±12.7 mm tolerance after the participant completed all corrections. This metric captures the combined outcome of violation detection and physical correction, reflecting the end-to-end inspection result that determines field acceptance. Table 5 shows the accuracy of inspection in two methods. Mean accuracy has mean difference of 5.1%. A Wilcoxon signed-rank test indicated no statistically significant difference between conditions (p = 0.204), with a small effect size (r = 0.32). The proportion of participants achieving ≥80% accuracy was higher in the AR condition (26/30, 87%) compared to Traditional (22/30, 73%). These results suggest that although AR did not produce a statistically significant improvement in overall accuracy, it may still offer practical benefits by helping more participants reach higher accuracy levels during inspection tasks.

Table 5. Accuracy comparison between inspection conditions

| Metric | Traditional (Mean ± SD) | AR (Mean ± SD) | Test | p-value | Effect Size (d) | Significance |
|---|---|---|---|---|---|---|
| Accuracy (%) | 81.7 ± 21.2 | 86.8 ± 19.3 | Wilcoxon | 0.204 | 0.32 | ns |
| Participants ≥ 80% | 22/30 (73%) | 26/30 (87%) | — | — | — | — |

In this study, accuracy was measured as a post-correction compliance rate, therefore it does not separately quantify detection sensitivity (the proportion of true violations identified) and correction precision (the positional accuracy of physical rebar adjustments).

## 5.2. Descriptive Statistics of Motion Capture Metrics

Table 6 presents the descriptive statistics for all motion capture metrics across both inspection conditions.

Table 6. Descriptive statistics of motion capture metrics (AR vs Traditional)

| Metric | Traditional (Mean ± SD) | AR (Mean ± SD) | Change (%) |
|---|---|---|---|
| Task Duration (s) | 440.1 ± 210.2 | 142.3 ± 56.5 | −67.7% |
| Walking Distance (m) | 43.2 ± 21.9 | 20.7 ± 9.0 | −52.0% |
| Trunk Flexion (°) | 53.3 ± 17.4 | 36.9 ± 20.7 | −30.8% |
| Neck Flexion (°) | 65.6 ± 19.6 | 44.1 ± 26.4 | −32.8% |
| Total Hand Path (m) | 128.2 ± 67.0 | 48.5 ± 16.8 | −62.2% |

AR-assisted inspection reduced task duration by 67.7% and walking distance by 52.0%, with corresponding reductions in total hand path length (62.2%). The variability across participants was notably lower under the AR condition for most metrics, particularly task duration (SD = 56.5 s vs. 210.2 s), indicating that the guided workflow produced more consistent inspection behavior regardless of individual experience level.

Postural analysis showed that reductions in high-risk posture exposure were disproportionately larger than reductions in mean joint angles. Mean trunk flexion decreased by 30.8%. A similar pattern was observed for the neck with 32.8% reduction in mean flexion angle. The mean neck flexion in the traditional condition (65.6°) exceeded the 25° threshold associated with increased cervical pain risk in occupational settings, and AR helped bringing this exposure closer to acceptable levels.

## 5.3. Task Performance and Movement Efficiency

AR-assisted inspection fundamentally changed task execution patterns compared to traditional methods (Table 7, Figure 6). Effect sizes for efficiency metrics indicated large effects, with task duration showing the strongest improvement. The magnitude of time reduction suggests AR helped decrease inefficiencies in manual inspection workflows specifically, the iterative process of consulting drawings, locating reference points, performing measurements, and recording findings. Movement pattern analysis showed that AR enabled more targeted inspection strategies: both walking distance and hand path reductions exceeded 50%, indicating participants navigated directly to violation locations rather than conducting exhaustive typical manual inspection.

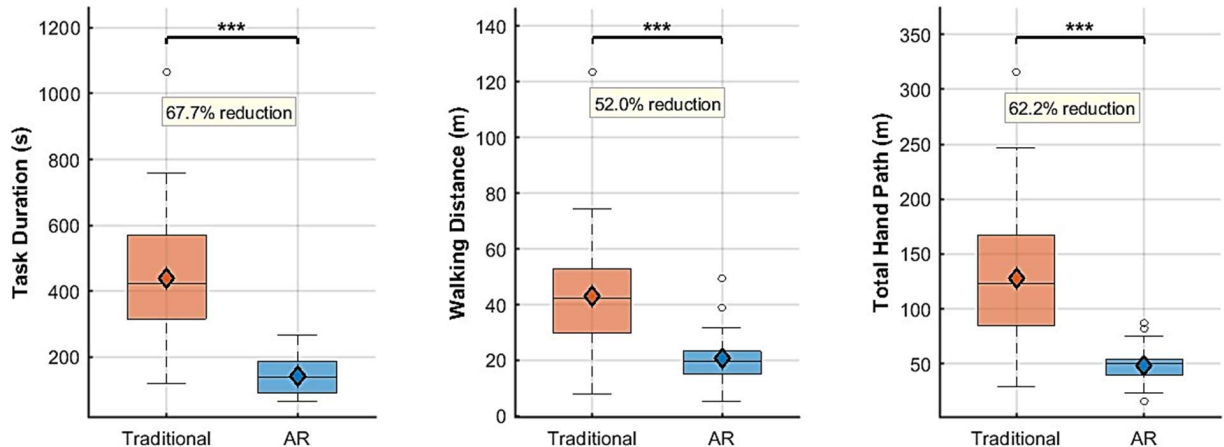

Figure 6. Task efficiency comparison in two methods

Table 7. Task performance efficiency metrics

| Metric | Test | p-value | Effect size (d) | 95% CI (Difference) | Significance |
|---|---|---|---|---|---|
| Task Duration | t-test | <0.001 | −1.488 | [-372.5, -223.0] | *** |
| Walking Distance | Wilcoxon | <0.001 | −0.768 | [-31.1, -13.9] | *** |
| Total Hand Path | t-test | <0.001 | −1.186 | [-104.9, -54.6] | *** |
| Significance levels: *** p < 0.001, ** p < 0.01, * p < 0.05, ns = not significant. Negative effect sizes indicate the AR condition scored lower than the Traditional condition. ||||||

*5.4. Postural Analysis: Trunk and Neck Kinematics*

Postural analysis demonstrated that AR-assisted inspection shifted workers away from biomechanically unsafe postures toward more neutral positions (Table 8, Figure 7).

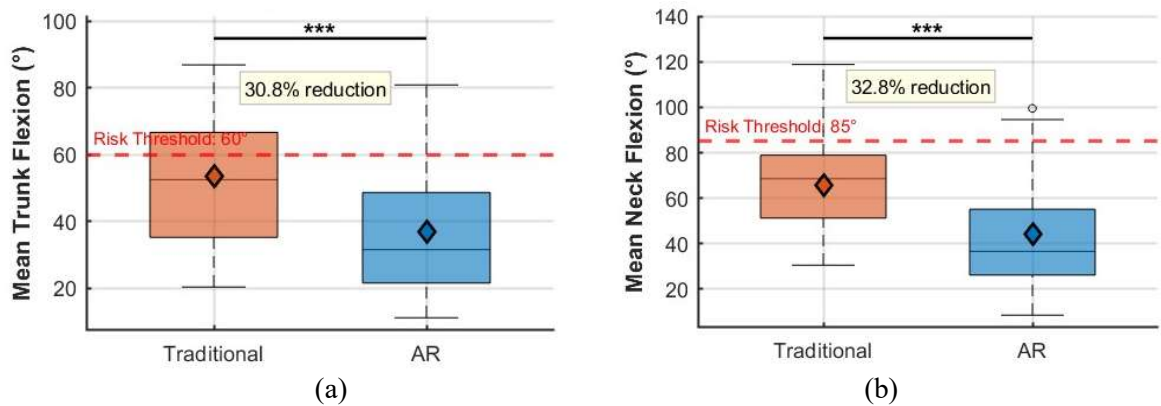

(a) (b)
Figure 7. Posture analysis; (a) Trunk, and (b) Neck

Mean trunk flexion reduction (30.8%) indicates a general shift toward upright postures. The medium-to-large effect sizes across trunk and neck metrics indicate these are not marginal improvements but represent meaningful ergonomic redesign of the inspection workflow. Neck posture also showed 32.8% mean angle reduction, indicating AR reduces exposure to positions associated with cervical strain.

Table 8. Task postural metrics (Trunk and Neck)

| Metric | Test | p-value | Effect size | 95% CI (Difference) | Significance |
|---|---|---|---|---|---|
| Trunk Flexion (Mean) | t-test | 0.0002 | −0.788 | [-24.2, -8.7] | *** |
| Neck Flexion (Mean) | t-test | 0.0005 | −0.710 | [-32.8, -10.2] | *** |
| Significance levels: *** p < 0.001, ** p < 0.01, * p < 0.05, ns = not significant. Negative effect sizes indicate the AR condition scored lower than the traditional condition. ||||||

ISO 11226 specifies time-dependent MAHT limits for postural risk assessment where sustained postures in the conditional zone become unacceptable if held beyond angle-specific thresholds, however, the brief nature of the inspection tasks resulted in no continuous postural segments exceeding these limits. The reported zone distributions shown in Figure 8 therefore represent the percentage of task time spent at each angular range (acceptable: trunk ≤20°, neck ≤25°; conditional: trunk 20-60°, neck 25-85°; not recommended: trunk >60°, neck ≥85°) based on instantaneous angular position, with the time-dependent criteria serving as a verification that no ergonomic violations occurred despite participants adopting conditional-zone postures.

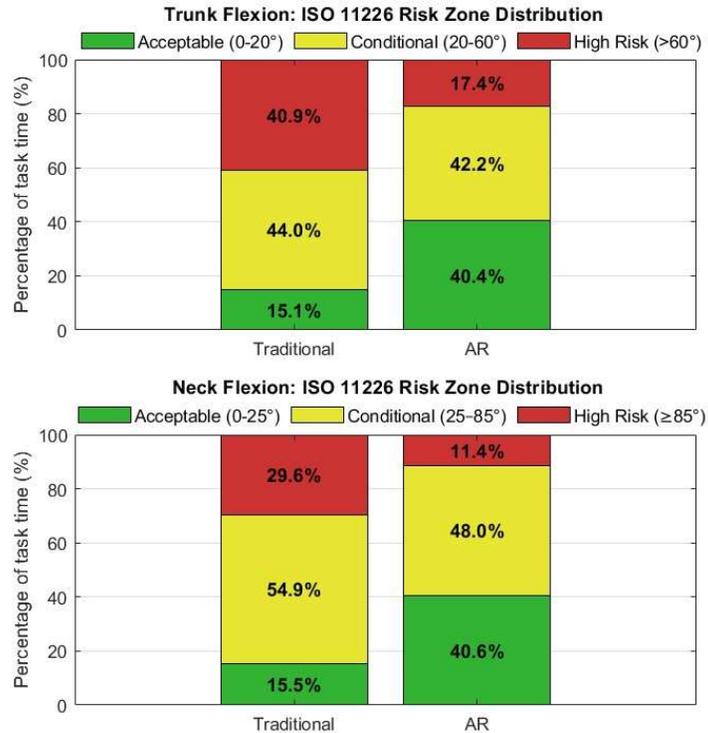

Figure 8. Distribution of trunk and neck time spent in three risk zones

The AR-assisted inspection method reduced the time workers spent in harmful postures compared to the traditional approach. For trunk flexion, high-risk posture time dropped from 40.9% to 17.4%, while acceptable posture time increased from 15.1% to 40.4%. Similar improvements occurred for neck flexion, where high-risk time decreased from 29.6% to 11.4%, and acceptable posture time rose from 15.5% to 40.6%. These results indicate that AR guidance allows inspectors to maintain healthier body positions throughout the rebar inspection task, because they no longer need to repeatedly bend forward to consult paper documents or take manual measurements, helping to reduce the risk of developing MSD during prolonged inspection activities.

### 5.5. NASA Task Load Index Analysis

The NASA-TLX analysis indicates that the AR system substantially reduced perceived workload across most task demand dimensions (Table 9, Figure 9).

Table 9. NASA-TLX dimensional-level analysis

| Dimension | Traditional (Mean ± SD) | AR (Mean ± SD) | p-value | Effect Size | Significance |
|---|---|---|---|---|---|
| Mental Demand | 55.5 ± 27.9 | 33.7 ± 21.7 | 0.0015 | −0.638 | ** |
| Physical Demand | 77.7 ± 23.6 | 18.8 ± 17.9 | <0.001 | −2.214 | *** |
| Temporal Demand | 52.5 ± 28.3 | 27.5 ± 21.4 | 0.0002 | −0.680 | *** |
| Performance | 53.5 ± 40.6 | 57.2 ± 40.2 | 0.5361 | +0.113 | ns |
| Effort | 69.2 ± 21.4 | 35.0 ± 21.1 | <0.001 | −0.750 | *** |
| Frustration Level | 56.3 ± 31.7 | 26.2 ± 22.9 | 0.0002 | −0.782 | *** |
| Significance levels: *** $p < 0.001$, ** $p < 0.01$, * $p < 0.05$, ns = not significant. Negative effect sizes indicate the AR condition scored lower than the Traditional condition. | | | | | |

The largest improvement occurred in Physical Demand, which decreased from 77.7 ± 23.6 in the traditional condition to 18.8 ± 17.9 with AR, corresponding to a very large effect (d=−2.214, p<0.001). This reduction suggests that the AR interface significantly reduces the physical burden associated with manual inspection activities, consistent with the objective motion capture results showing reduced trunk and neck flexion and shorter hand movement paths.

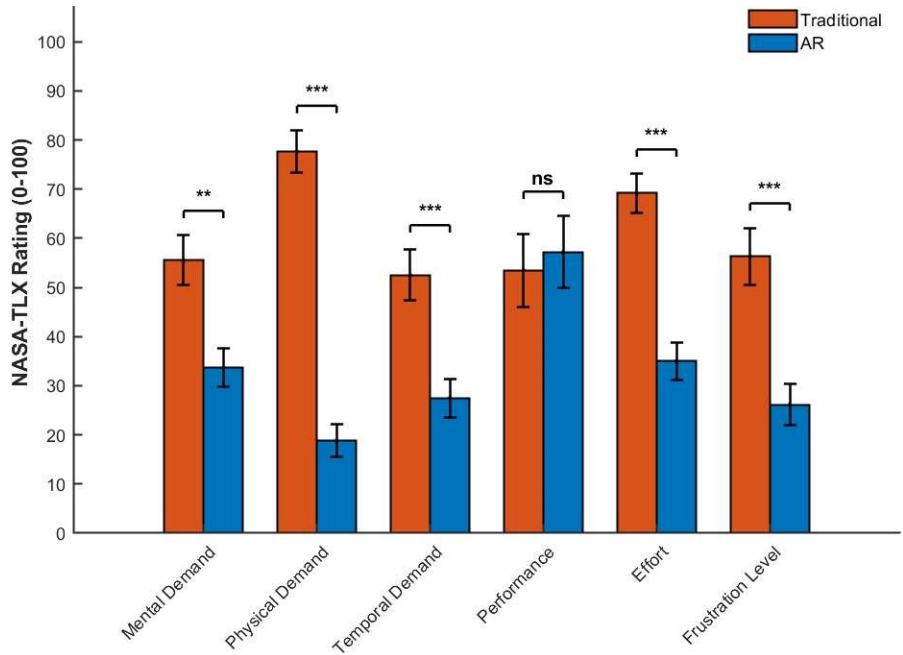

Figure 9. NASA-TLX score comparison in two different inspection methods

Significant reductions were also observed in Effort (d= −0.75, p<0.001), Frustration Level (d= −0.782, p= 0.0002), and Temporal Demand (d= −0.68, p = 0.0002), indicating that the AR system not only decreases physical strain but also improves the perceived ease and pace of task execution. Mental Demand was also significantly lower under the AR condition (d= −0.638, p=0.0015), although its effect size was smaller than those observed for physical and effort-related dimensions, suggesting that the primary benefits of AR are ergonomic rather than purely cognitive. In contrast, Performance did not differ significantly between conditions (d= 0.113, p=0.536), indicating that the reductions in workload were not associated with any perceived decline in inspection quality. This dissociation between reduced workload and stable performance perception suggests that AR improves task efficiency by simplifying information access and reducing unnecessary physical interactions. Consistent with these dimension-level findings, the overall NASA-TLX score decreased from 60.8 ± 17.7 to 33.1 ± 14.5, representing a 45.6% reduction in perceived workload. Together with the objective reductions in task duration and ergonomic risk, these results indicate that AR meaningfully improves both the physical and perceptual aspects of the inspection process.

### 5.6. System Usability Scale Results

SUS scores were interpreted using established benchmarks [52]: scores below 50 indicate "Not Acceptable" usability, 50-70 represent "Marginal" acceptability, and scores above 70 are considered "Acceptable." System usability assessment demonstrated above-average acceptance with notable individual variability as shown in Figure 10. The mean SUS score of 76.1 places the system in the "B-C" grade range which is a level typically associated with commercially viable products rather than research prototypes. The distribution characteristics provide additional context: 83% acceptability rate (scores ≥70) indicates broad user acceptance, while the 35.0–97.5 range shows that experiences varied from marginal to excellent. The median (78.8) exceeding the mean (76.1) indicates the distribution is slightly left-skewed, with the low outlier pulling the mean downward. From a technology adoption perspective, the critical finding is that 54% of users rated the system as "good" (>75) and 83% as "acceptable" despite minimal training and no prior AR experience. This usability performance is particularly notable given participants' limited AR familiarity (mean=2.8/5), indicating that the learning curve is manageable for most users and that the interface successfully accommodates novice users.

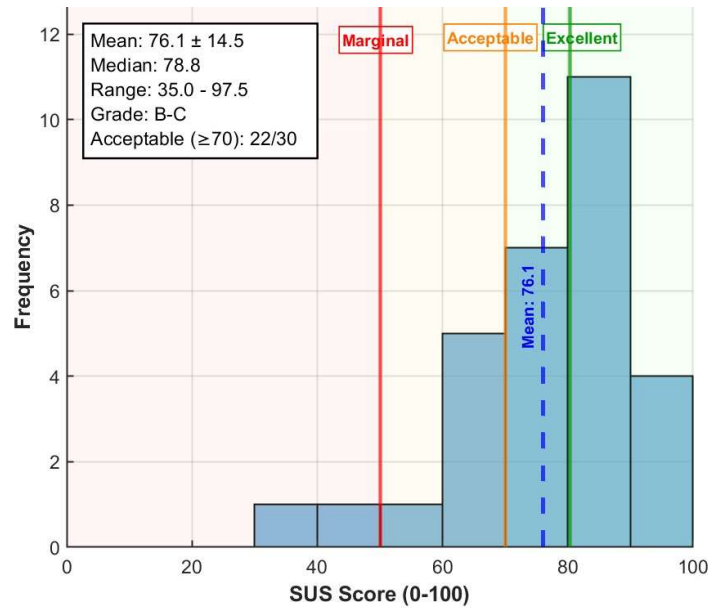

Figure 10. SUS score frequency plot

## 6. Discussion

This study provided empirical evidence that AR-assisted rebar inspection substantially reduces ergonomic stressors compared to traditional manual methods. The 30.8% reduction in mean trunk flexion represents biomechanically significant improvements with direct implications for MSD prevention. The reduction in high-risk flexion exposure therefore translates to substantially reduce cumulative spinal loading during inspection work. Given that construction workers experience MSDs at rates 40% higher than the all-industry average [35], with back injuries representing a leading cause of lost work time [37], interventions that help reducing biomechanical risk factors address a critical industry need. Multiple factors contribute to the postural improvements observed with AR. First, AR visualization eliminates the need to repeatedly check paper drawings held at waist level, which typically require sustained trunk flexion. Second, automated violation highlighting removes the requirement for close-range visual inspection and manual measurements that necessitate deep bending. Third, the hands-free nature of the head-mounted display allows inspectors to maintain more upright postures while accessing information. The reduction in neck flexion and neck high-risk time are particularly important, as sustained neck flexion beyond 85° has been associated with increased neck pain risk in occupational settings [62].

The 67.7% reduction in task completion time combined with 43.8% decrease in mental demand ratings indicates AR provides dual benefits: improved efficiency and reduced cognitive load. These findings challenge assumptions that technology-mediated speed improvements necessarily increase cognitive stress. Instead, the data suggests AR's intuitive visualization reduces mental effort required to transform 2D design representations into 3D spatial understanding. Movement pattern analysis provides insight into efficiency mechanisms: the 52.0% reduction in walking distance and 62.2% decrease in hand path indicates AR guidance enables more direct navigation to problem areas rather than exhaustive sequential surveys. The maintained Performance self-ratings across conditions (no significant difference, p = 0.536), supported by objective accuracy data showing no significant difference between conditions, confirm that time savings did not compromise inspection quality.

The SUS score of 76.1 (grade B-C, "above average") represented solid usability performance, particularly considering participants' limited prior AR experience and brief training period. This finding is encouraging for real-world deployment potential. The 83% acceptability rate (scores ≥70) suggests the learning curve is manageable. However, the range of scores (35.0–97.5) reveals individual variability, with one participant rating the system as unacceptable. The lack of correlation between inspection experience and SUS scores (r=0.18, p=0.31) suggests the interface's usability exceeds domain-specific knowledge, potentially lowering training barriers compared to traditional inspection requiring extensive drawing interpretation expertise.

### 6.1. Practical Implications for Construction Practice

The demonstrated benefits have implications for construction practice. If the 70.7% reduction in extreme trunk flexion observed in this controlled study translates to field settings, it could substantially decrease low back injury incidence and severity among inspection personnel. The 67.7% time savings could contribute to shorter project durations and reduced costs, as quality control inspections represent critical path activities. The improved digital documentation capabilities (automated violation logging, photographic evidence) could streamline communication between inspectors, contractors, and project managers. However, successful field implementation requires addressing practical challenges not captured in laboratory settings. Real construction sites present environmental challenges including dust, lighting variation, outdoor conditions, and congested workspaces that may affect AR tracking accuracy and display visibility.

### 6.2. Limitations and Future Research

The controlled laboratory environment differs from actual construction sites. The standardized rebar assembly lacked the complexity, congestion, and accessibility constraints of real structures. Environmental factors (poor lighting, outdoor conditions, dust, noise) were absent. Task durations (mean <10 minutes) were substantially shorter than typical inspection shifts, excluding assessment of sustained effects or fatigue-related changes. These factors limit direct generalizability of quantitative results, though fundamental ergonomic principles should remain valid. Participants were mostly young adults (mean age 28.8 years) with limited professional experience (79% had no inspection experience). While within-subjects design controlled for individual differences, the sample may not represent the full construction workforce demographic range. Older workers or those with extensive traditional inspection experience might demonstrate different adaptation patterns or technology acceptance. This study evaluated single-session effects. Longer-term considerations including learning curves, sustained performance over work shifts, visual fatigue, and potential HMD-related neck strain from prolonged wear were not assessed. The HL2 mass (~566 g) could introduce cervical loading concerns during extended use, though our brief sessions showed no adverse effects.

Future research should address these limitations through longitudinal field studies on active construction projects, enabling evaluation under realistic environmental conditions and integration with actual project workflows. Investigation of different rebar configurations (walls, columns, complex geometries) would test system generalizability. Extended-use studies (8-hour shifts over multiple days) are needed to assess fatigue effects, learning curves, and sustained ergonomic benefits.

### 7. Conclusion

This study addressed a gap in the AR-construction literature by providing simultaneous objective biomechanical and validated subjective evidence for the human factors impact of AR-assisted rebar inspection. The convergence of motion capture data with NASA-TLX and SUS results across 30 participants produced a consistent picture: AR-assisted inspection shifted workers away from high-risk postures, reduced task duration and movement demands, lowered perceived workload across nearly all dimensions, and achieved above-average usability without compromising inspection accuracy. Two findings deserve particular emphasis. First, the reductions in extreme posture exposure were larger than the reductions in mean joint angles, indicating that AR selectively eliminates the most hazardous subtasks rather than producing a uniform postural shift. This pattern has direct implications for musculoskeletal injury prevention, as cumulative high-risk exposure is a stronger predictor of injury than average posture. Second, the subjective and objective data aligned closely confirming that the largest perceived workload reduction occurred in physical demand, consistent with the measured postural improvements. These results should be interpreted within the constraints of a controlled laboratory study. The standardized rebar assembly, short task durations, and predominantly young participant sample limit direct generalizability to field conditions. Translation to practice will require field validation under realistic environmental conditions, extended-use studies assessing fatigue and sustained performance, and evaluation across diverse rebar configurations and workforce demographics. This multi-domain evidence base study established a foundation for evidence-based AR adoption decisions in construction quality control.

**Acknowledgement**

The authors appreciate the support of the Federal Railway Administration (FRA) BAA project number 693JJ621C000010; the National Science Foundation, Division of Information & Intelligent Systems (IIS), CISE, Hardening the Data Revolution DSC (Grant/Award No. 2123346); the U.S. Department of Transportation (Award No. 69A3552348306) through the USDOT Southern Plains Transportation Center (SPTC); the Office of Naval Research, Structural Reliability ONR 331 (Grant No. 13620990, Award No. N00014-22-1-2638), the Office of Naval Research, Structural Reliability ONR 331 (Grant No. 13106652, Award No. N00014-21-1-2929), and National Aeronautics and Space Administration (NASA) Cooperative Agreement EPSCoR Research Infrastructure Development (RID) Grant 80NSSC22M0044, through the New Mexico Space Grant Consortium Grant 2RAFY. The authors also acknowledge the support from the Department of Civil, Construction, and Environmental Engineering of the University of New Mexico (UNM); and the Center of Advanced Research and Computing (CARC) at UNM. The authors thank Michael Carl and Hayden Lammons for assistance with experiments.